# Non-local effects on the heavy-ion fusion at sub-barrier energies

L. C. Chamon, M. S. Hussein
*Instituto de Física da Universidade de São Paulo,*
*Caixa Postal 66318, 05315-970, São Paulo, SP, Brazil.*

L. F. Canto
*Universidade Federal do Rio de Janeiro, RJ, Brazil.*
(Dated: December 2, 2018)

We investigate the effect of Pauli non-locality in the heavy-ion optical potential on sub-barrier fusion reactions. The São Paulo potential, which takes into account the Pauli non-locality and has been widely used in analyzing elastic scattering, has also recently been applied to heavy-ion fusion. However, the approximation employed in deriving the São Paulo potential, based on the Perey-Buck semi-classical treatment of neutron induced reactions, must be assessed for charged particles tunneling through a barrier. It is the purpose of this note to look into this question. We consider the widely studied system $^{16}$O + $^{208}$Pb at energies that span the barrier region from 10 MeV below to 10 MeV above. It seems that the non-locality plays a minor role. We find the São Paulo potential to be quite adequate throughout the region.

PACS numbers: 25.60.Pj

## I. INTRODUCTION

Heavy ion reactions in the vicinity of the Coulomb barrier have revealed many surprising features over the last two decades. The enhancement of the measured fusion cross section when compared to the simple, one-dimensional barrier penetration model [BPM] was eventually attributed to the influence of the couplings to several reaction channels [1]. More recently, when comparing the optical potentials, taken to be of the Woods-Saxon (WS) form, that describe the elastic scattering angular distribution to the ones that describe fusion, it was found that the diffuseness of the real part of the latter is almost twice that of the former [2, 3]. Quite recently, another interesting feature in fusion was discovered. The data at deep sub-barrier energies were found to be hindered when compared to theoretical cross sections with potentials adjusted to fit the data at higher energies [4]. This latter phenomenon is suggested to be a consequence of internal repulsion that makes the potential well shallower [5, 6]. Such repulsion may come about due to the operation of Pauli blocking, not considered in the usual double-folding potentials which are used to generate the WS ones alluded to above. What about the effect of exchange? Here we give the answer.

Both channel coupling and exchange leads to non-locality in the effective potential. In [7] we called the two type of non-localities, the Feshbach and Pauli non-localities respectively. The Pauli non-locality is present in the bare potential and has a very mild energy dependence. On the other hand, the Feshbach non-locality arising from channel couplings is accompanied by a rather strong energy dependence, which conspicuously manifests itself in the form of the so-called Threshold Anomaly (TA) arising from the dispersion relation manifestly obeyed by the Feshbach, polarization, potential (the TA is an unfortunate name since the absence of the anomaly is in fact THE anomaly). Of course, when used in coupled-channel calculations, the non-locality is transformed into a non-dispersive energy dependence. In so far as the Pauli non-locality is concerned, the resulting locally equivalent non-dispersive energy dependent potential has been constructed by our group [7, 8, 9, 10] and it is coined the São Paulo potential. We shall use this potential here to investigate the relevance of the Pauli, exchange, non-locality on the fusion cross section at energies around the barrier. The paper is organized as follows. In section II we give an account of the São Paulo potential. In Section III we calculate the fusion cross section for $^{16}$O + $^{208}$Pb and compare it with the data. We assess the effect of the Pauli non-locality and found it to be minor. In Section IV we present our concluding remarks.

## II. THE SÃO PAULO OPTICAL POTENTIAL

The São Paulo potential (SPP) has been in use for about a decade especially in the analysis of heavy-ion elastic scattering data [8, 10, 11, 12, 13, 14, 15, 16, 17, 18, 19, 20, 21, 22, 23, 24, 25]. It is based in part on the idea of single folding employed for alpha-nucleus scattering by Jackson and Johnson [26] in order to get the Perey-Buck (PB) nonlocal effects [27] extended to heavy ions. In Refs. [7, 8, 9], the Jackson-Johnson idea was extended to heavier systems and the resulting potential, which is a modification of the double-folding one, was shown to be quite accurate in accounting for a large body of elastic scattering data for a wide range of systems, including $^{16}$O + $^{208}$Pb. The PB non-locality was originally derived for neutron-nucleus scattering. The effect of the Coulomb repulsion seems to be as well described accurately by the semi-classical PB-based SPP.

Recently, the SPP has been employed to the calcula-tion of heavy-ion fusion cross sections [28]. Here, one would expect more sensitivity to the tunneling effect and consequently on the sensitivity of the PB non-locality in the SSP on this important quantal phenomenon. In this paper, we discuss this issue by a careful analysis of the non-locality in the heavy-ion fusion at below- and above-barrier energies.

The heavy-ion optical potential is non-local owing to two effects: the dispersive non-locality related to channel couplings (the Feshbach non-locality), and to the Fermionic nature of the constituents (exchange or Pauli non-locality). The Feshbach non-locality is accompanied by strong energy-dependence which manifests itself through the dispersion relation satisfied by the real and imaginary parts of the Feshbach channel coupling polarization interaction. The Pauli non-locality, on the other hand, is accompanied by a weak energy dependence arising from the energy content of the effective nucleon-nucleon interaction (the G-matrix). In this work we are interested only in the effect of the Pauli non-locality. Thus we assume a local energy-independent imaginary potential of the total optical potential and a corresponding local electromagnetic interaction. We can write [8, 9]

$$U(\vec{R}, \vec{R}'; E) = V(\vec{R}, \vec{R}'; E) + [iW(R') + V_C(R')]\delta(\vec{R} - \vec{R}') \quad (1)$$

where $V$ denotes the real part of the potential, $W$ the imaginary part and $V_C$ is the Coulomb potential. The scattering integro-differential Schroedinger equation that has to be solved is then,

$$-\frac{\hbar^2}{2\mu}\nabla^2 \psi(\vec{R}) + \int U(\vec{R}, \vec{R}'; E)\psi(\vec{R}')d\vec{R}' = E\psi(\vec{R}). \quad (2)$$

Perey and Buck [27] and Frahn and Lemmer (FL) [29] suggested the following simple Gaussian for $V(\vec{R}, \vec{R}'; E)$, after ignoring its energy dependence,

$$V(\vec{R}, \vec{R}') = V_{NL}\left(\frac{R+R'}{2}\right)\frac{1}{\pi^{3/2}b^3}e^{-|\vec{R}-\vec{R}'|^2/b^2}, \quad (3)$$

where $b$ is the range of the Pauli non-locality. It is convenient to write down the usual expansion in partial waves,

$$\psi(\vec{R}) = \sum i^\ell (2\ell+1)\frac{u_\ell(R)}{kR}P_\ell[\cos(\theta)], \quad (4)$$

$$V(\vec{R}, \vec{R}') = \sum \frac{2\ell+1}{4\pi RR'}V_\ell(R, R')P_\ell[\cos(\phi)], \quad (5)$$

$$V_\ell(R, R') = V_{NL}\left(\frac{R+R'}{2}\right)\frac{1}{\pi^{1/2}b} \times \left[Q_\ell\left(\frac{2RR'}{b^2}\right)e^{-\left(\frac{R-R'}{b}\right)^2}(-)^{\ell+1} Q_\ell\left(\frac{-2RR'}{b^2}\right)e^{-\left(\frac{R+R'}{b}\right)^2}\right], \quad (6)$$

where $Q_\ell$ are polynomials and $\phi$ is the angle between $\vec{R}$ and $\vec{R}'$ [27]. Thus, the integro-differential equation can be recast into the following form:

$$-\frac{\hbar^2}{2\mu}\frac{d^2u_\ell}{dR^2} + [E - V_C(R) - iW(R) - \frac{\ell(\ell+1)\hbar^2}{2\mu R^2}]u_\ell(R) = \int_0^\infty V_\ell(R, R')u_\ell(R')dR'. \quad (7)$$

The local-equivalent $\ell$- and energy-dependent potential is defined as

$$V_{LE}(R) + iW_{LE}(R) = \frac{1}{u_\ell(R)}\int_0^\infty V_\ell(R, R')u_\ell(R')dR'. \quad (8)$$

The $\ell$-dependence of the interaction $V_{LE}(R)$ is in fact negligible and it approximately satisfies the following non-linear equation:

$$V_{LE}(R) = V_{NL}(R)e^{-\gamma[E - V_C(R) - V_{LE}(R) - iW(R)]}. \quad (9)$$

In accord with Jackson and Johnson [26], the non-locality parameter is given by:

$$\gamma = \mu b^2/2\hbar^2. \quad (10)$$

Here, $\mu$ is the reduced mass of the two nuclei and $b$ is related to the nucleon-nucleus non-locality parameter $b_0$, determined through systematics to be 0.85 fm [27], through $b = b_0 m_0/\mu$, with $m_0$ being the nucleon mass.

For neutron-nucleus systems, Perey and Buck have associated the nonlocal interaction with a Woods-Saxon shape potential. For nucleus-nucleus, we have associated [8, 9] the energy-independent nonlocal real potential $V_{NL}[(R+R')/2)]$, that appears in the PB and FL form, with the double folding potential,

$$V_{NL}(R) = V_{Fold}(R), \quad (11)$$

where $V_{Fold}(R)$ is calculated following the procedure described in Ref. [9]. With this, and ignoring $W(R)$ in Eq. (9), which is of minor consequence, we obtain the SPP, namely,

$$V_{LE}(R) = V_{Fold}(R)e^{-\gamma[E - V_C(R) - V_{LE}(R)]}. \quad (12)$$

Eq. (12) has been commonly expressed by the equivalent form in terms of the local relative velocity $v^2(R) = 2[E - V_{LE}(R) - V_C(R)]/\mu$, and the SSP acquires the simple form,

$$V_{LE}(R; E) = V_{Fold}(R)e^{-4v^2(R)/c^2} \quad (13)$$

where $(b_0 m_0 c/\hbar)^2$ is numerically very close to 4.

### III. APPLICATION TO FUSION

To test the use of the SPP in tunneling problems and compare it to the exact solution of the integro-differential equation, we have considered the fusion system $^{16}$O + $^{208}$Pb at center of mass energies that span a region 10 MeV below the Coulomb barrier up to 10 MeV above it. In the analysis, we have assumed a Woods-Saxon shape inner imaginary potential to simulate the flux absorption by the fusion process, with the following parameters: $W_0 = 200$ MeV, $r_{i0} = 0.8$ fm, and $a_i = 0.2$ fm. In figure 1 we show the fusion cross section obtained through three different procedures: i) using the exact integro-differential equation, Eq. 2, ii) the SPP (Eq. 13) as the local-equivalent potential, and iii) the simple double-folding interaction as the local nuclear interaction (no non-locality). There is hardly any difference among the results of the three calculations, represented by the solid line in Fig. 1, indicating that non-local effects in fusion are negligible. The dashed line in Fig. 1 represents the coupled channel (CC) results with the SPP interaction. The CC calculation has been performed considering the couplings to inelastic channels as described in Ref. [28]. Clearly the CC effect (Feshbach non-locality) plays a much more important role than the Pauli non-locality.

To understand this behavior, in Fig. 2 we show the SPP, the folding potential, and the angular momentum dependence of the real part of the exact local equivalent potential obtained through Eq. 8. The exact $V_{LE}$ has insignificant $\ell$-dependence and it is very well described by the SPP except at very small distances ($R \approx 0$). The folding potential has a larger strength at small distances, of little consequence to the tunneling problem in this highly absorbing system. In fact, for energies near the barrier height and distances close to the barrier radius ($R_B$), the relative velocity is very low and therefore Eq. 13 indicates that the SPP is very similar to the folding potential at the surface region. This explain why the non-locality in the heavy-ion interaction has no relevance for near-barrier fusion, though very important in describing the elastic scattering data at higher energies.

## IV. CONCLUSION

In this paper, we have carefully assessed the effect of the Pauli non-locality, as treated in the São Paulo potential, on the heavy ion fusion at energies in the vicinity of the Coulomb barrier. We also calculated the effect using the original double-folding + Perey/Buck non-locality by solving the corresponding integro-differential Schrodinger equation and compared these calculations with the case with no non-locality (the local double-folding potential). In earlier works, we have demontrated that the Pauli non-locality is very important to describe in a consistent manner the elastic scattering process in a wide energy range. Here, it was found that exchange effects that give rise to the Pauli non-locality have a minor role on tunneling at energies around the barrier.

**Acknowledgments**

This work was partially supported by Fundação de Amparo à Pesquisa do Estado de São Paulo (FAPESP), Fundação de Amparo à Pesquisa do Estado do Rio de Janeiro (FAPERJ), and Conselho Nacional de Desenvolvimento Científico e Tecnológico (CNPq).

FIG. 2: The SPP, folding and real part of the exact local equivalent potential for three different partial waves, calculated for $E_{c.m.} = 69.6$ MeV. The position of the barrier radius is indicated in the figure.

FIG. 1: Fusion data (from Refs. [30, 31]) for the $^{16}$O + $^{208}$Pb system. The solid line represents the results of three different calculations as described in the text. The dashed line corresponds to CC calculations as presented in Ref. [28].

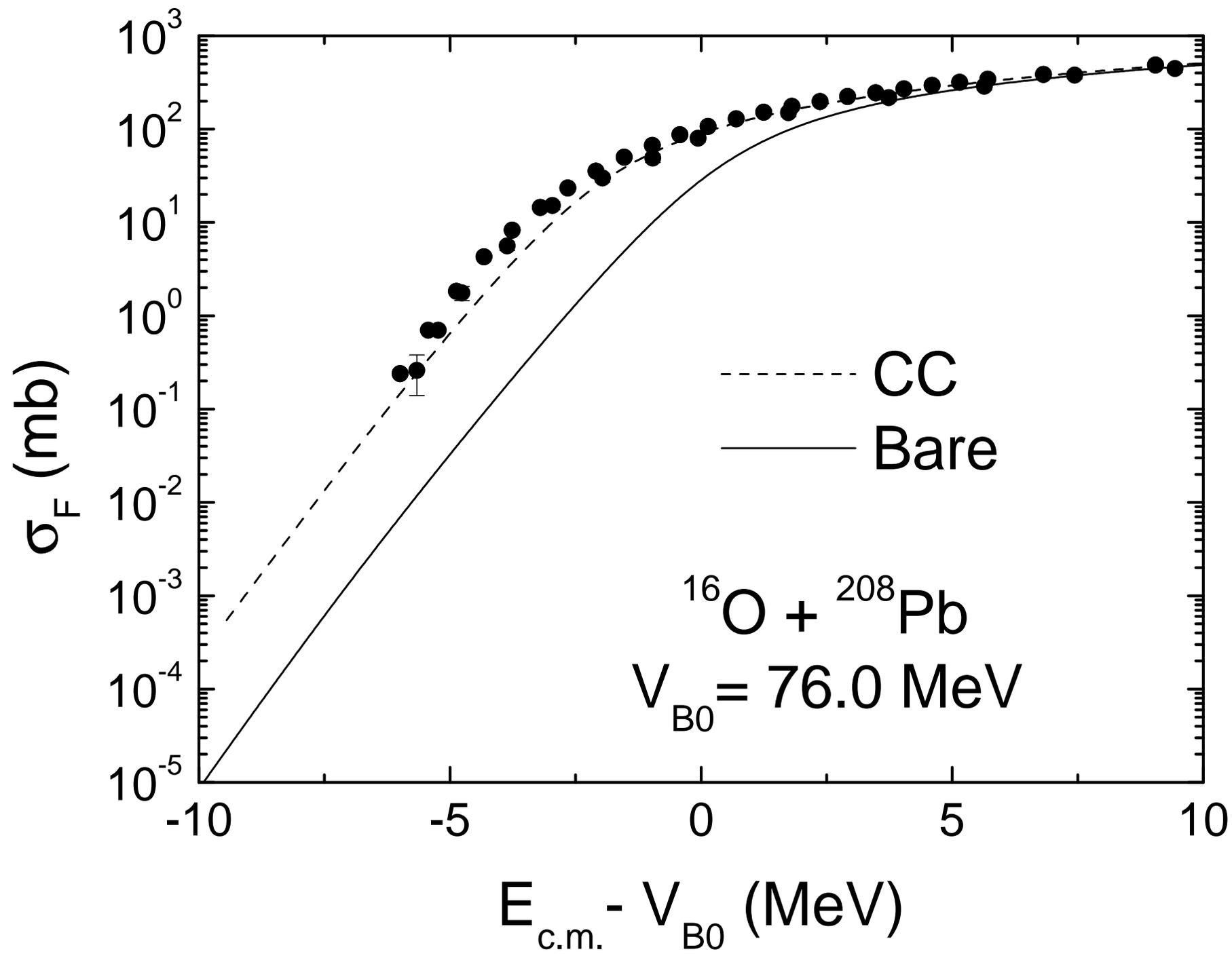

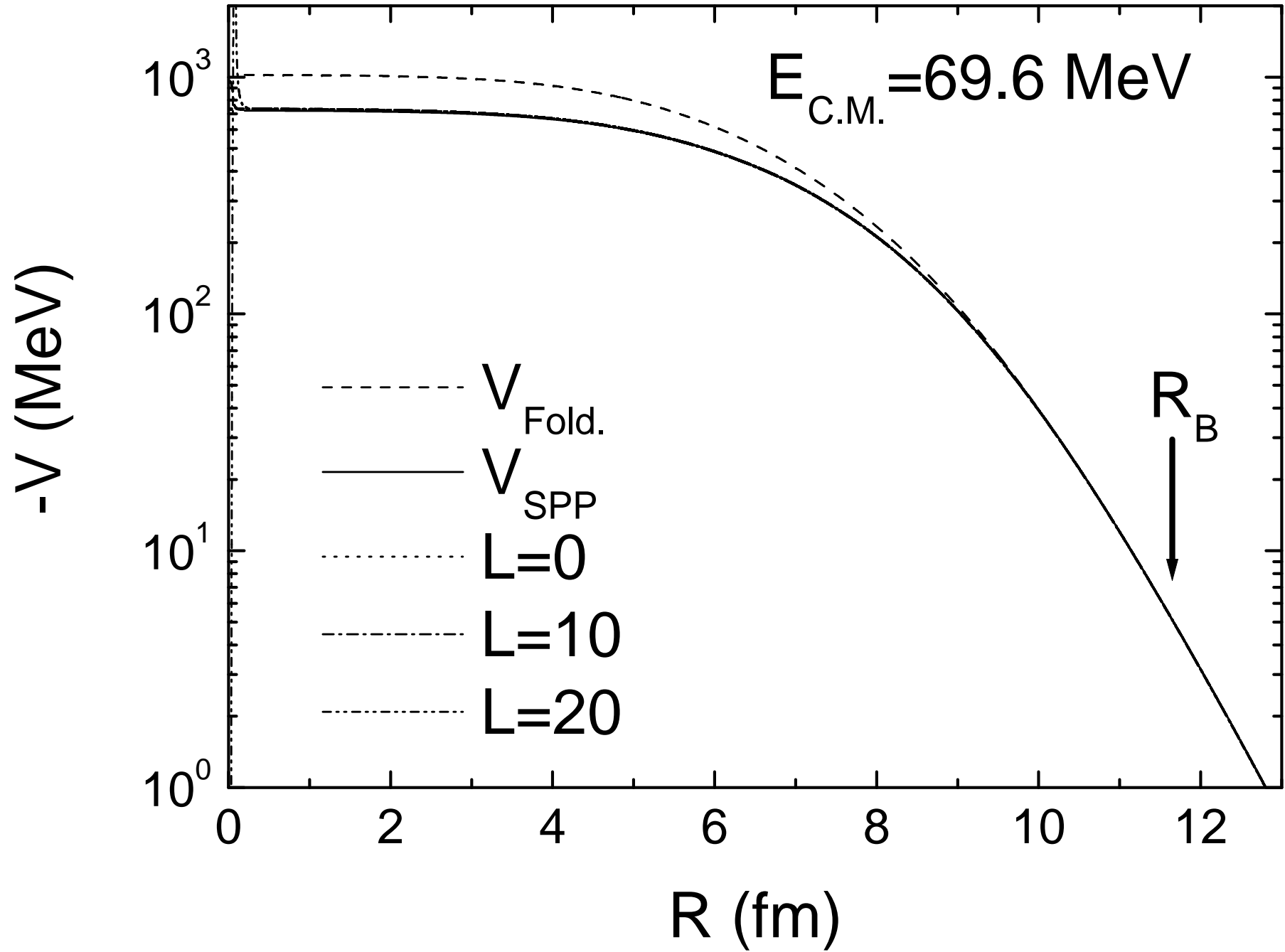